\documentclass[10pt]{iopart}
\usepackage{graphicx}
\usepackage{iopams}  
\begin{document}

\title{$J/\psi$ production at RHIC-PHENIX}
\author{Susumu X Oda (for the PHENIX Collaboration)}
\address{Center for Nuclear Study, Graduate School of Science, University of Tokyo, 7-3-1 Hongo, Bunkyo, Tokyo 113-0033, Japan}
\ead{oda@cns.s.u-tokyo.ac.jp}

\begin{abstract}
The $J/\psi$ is considered to be among the most important probes for the deconfined quark gluon plasma (QGP) created by relativistic heavy ion collisions. 
While the $J/\psi$ is thought to dissociate in the QGP by Debye color screening, there are competing effects from cold nuclear matter (CNM), feed-downs from excited charmonia ($\chi_c$ and $\psi'$) and bottom quarks, and regeneration from uncorrelated charm quarks. 
Measurements that can provide information to disentangle these effects are presented in this paper. 
\end{abstract}

\section{Introduction} 
The yield of heavy quarkonia is expected to be suppressed in the QGP due to the Debye screening of the color charge~\cite{Matsui_Satz}. 
The $J/\psi$ is especially promising because of its large production cross section and di-lepton decay channels, which make it easily detected. 
The PHENIX experiment at RHIC is able to detect the $J/\psi$ at midrapidity ($|y|<0.35$) via its decay to $e^+e^-$ and at forward rapidity ($1.2<|y|<2.2$) via its $\mu^+\mu^-$ decay. 
Models of $J/\psi$ production in heavy ion collisions at RHIC energy contain a number of important competing effects, including modification of the $J/\psi$ yield by the CNM effects, destruction of $J/\psi$ due to interactions with thermal gluons in the QGP, reduced feed-down from excited charmonium states that melt just above the QGP transition temperature, bottom quark decay and enhancement of the yield due to coalescence of uncorrelated charm pairs~\cite{Satz}. 
The PHENIX Au+Au data at $\sqrt{s_{NN}}=200$~GeV showed that $J/\psi$ suppression at forward rapidity is larger than that at midrapidity and the suppression at midrapidity is similar to that observed by NA50 at SPS in Pb+Pb collisions at $\sqrt{s_{NN}}=17.3$~GeV~\cite{Adare2}. 
However, these results are not well understood theoretically. 
Systematic study of $J/\psi$ production in heavy ion collisions across the entire range of $N_{part}$ is needed to disentangle the competing effects. 
PHENIX recorded Cu+Cu collisions at $\sqrt{s_{NN}}=200$~GeV to obtain precise data in the range $N_{part}\le 126$, where Au+Au data is limited by statistics and systematic uncertainty and the CNM effects might be dominant~\cite{Adare4}. 
Feed-downs into $J/\psi$ in $p+p$ collisions were measured at the same energy and are important in the picture of the sequential dissociation of quarkonia by different binding energy. 
Measurements of $J/\psi$ in Cu+Cu and Au+Au collisions and feed-downs into $J/\psi$ in $p+p$ collisions are presented in this paper. 
The elliptic flow of $J/\psi$ can set a constraint on coalescence of charm quarks and the measurement is reported in~\cite{Silvestre}. 

\section{Measurement of feed-downs into $J/\psi$ in $p+p$ collisions}
An inclusive $J/\psi$ measurement in $p+p$ collisions at $\sqrt{s}$=200~GeV has been reported by PHENIX~\cite{Adare1}. 
Recently, PHENIX separately measured the feed-down contributions of $\chi_c$, $\psi'$ and bottom quarks to $J/\psi$ in $p+p$ collisions at midrapidity ($|y|<0.35$) at the same energy. 
The preliminary results are described in this section. 

\begin{figure}[htbp]
  \begin{center}
    \includegraphics[width=\linewidth]{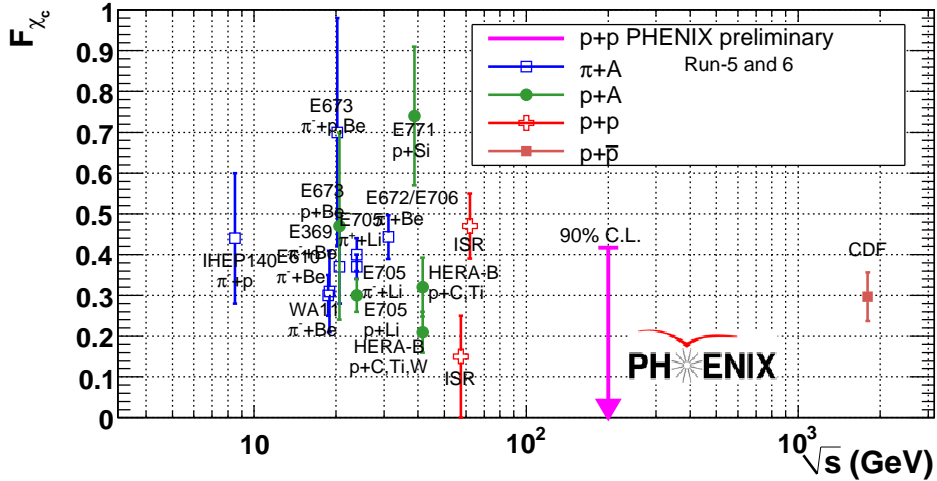}
    \caption{
      The fractions of $J/\psi$ from the $\chi_c$ decay, $F_{\chi_c}$, in various collision systems as a function of the center of mass energy, $\sqrt{s}$.   
      The 90\% confidence level upper limit of $F_{\chi_c}$ in $p+p$ collisions at $\sqrt{s}$=200~GeV which is the PHENIX preliminary result is represented by the arrow. 
      The experiment names are written in the figure. 
    }
    \label{fig1}
  \end{center}
\end{figure}

The fraction of $J/\psi$ from the $\chi_c$ decay, $F_{\chi_c}$, was measured via the decay chain of $\chi_c\rightarrow J/\psi+\gamma\rightarrow e^+e^-\gamma$ with $4.1\times 10^3$ reconstructed $J/\psi$. 
Branching ratios of $\chi_{cJ}$ to $J/\psi+\gamma$ are 1.3\% ($\chi_{c0}$), 36\% ($\chi_{c1}$) and 20\% ($\chi_{c2}$)~\cite{PDG}, and the contribution of $\chi_{c0}$ was neglected in the analysis. 
The 90\% confidence level upper limit obtained for $F_{\chi_c}$ is 0.42. 
Figure~\ref{fig1} shows the upper limit of $F_{\chi_c}$ with results from other experiments. 

The ratio of cross sections of $\psi'$ to $J/\psi$ was measured in the $e^+e^-$ decay mode and the feed-down fraction of $J/\psi$ from $\psi'$ decay is $0.086\pm 0.025$. 
The cross section of bottom quarks was measured by electron-hadron correlation~\cite{Morino} and the feed-down fraction of $J/\psi$ from decay of bottom and anti-bottom quarks is $0.036^{+0.025}_{-0.023}$ for $p_{T,J/\psi}>$0~GeV/$c$. 
Branching ratios of~\cite{PDG} were used to obtain these feed-down fractions. 
A set of theoretically predicted feed-down fractions~\cite{Vogt1} is almost consistent with the PHENIX preliminary results. 

\section{Measurement of $J/\psi$ in Cu+Cu and Au+Au collisions}
\begin{figure}[htbp]
  \begin{center}
    \includegraphics[width=\linewidth]{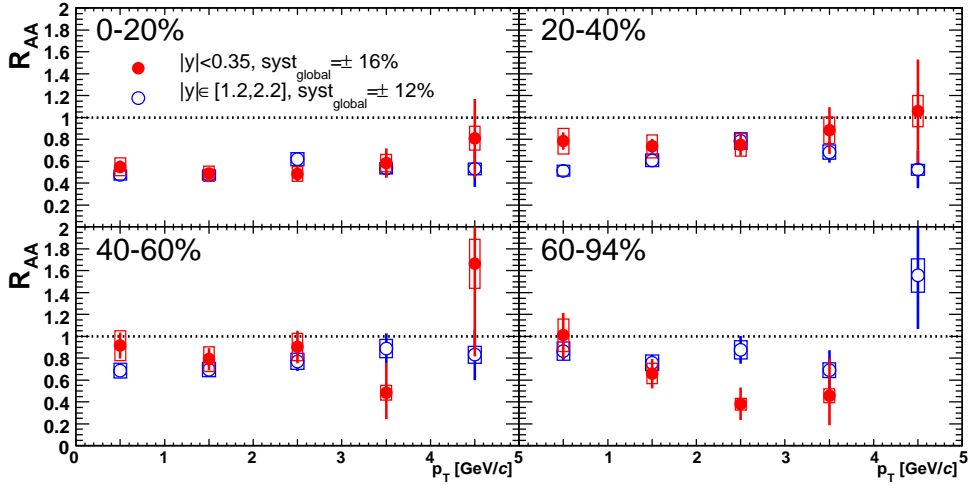}
    \caption{
      $R_{AA}$ vs transverse momentum $p_{T}$ for $J/\psi$ production in Cu+Cu collisions via $e^+e^-$ decay at midrapidity (closed circles) and $\mu^+\mu^-$ decay at forward rapidity (open circles). 
      Centrality bins are written in figures. 
      No strong $p_T$ dependence of $R_{AA}$ is observed in Cu+Cu collisions. 
    }
    \label{fig2}
  \end{center}
\end{figure}

\begin{figure}[htbp]
  \begin{center}
    \includegraphics[width=\linewidth]{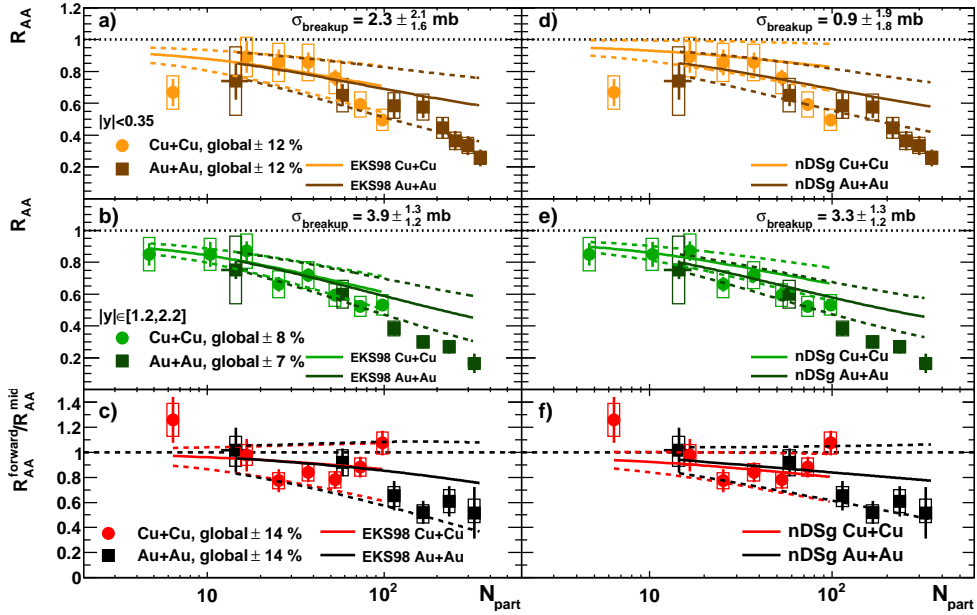}
    \caption{
      a) $R_{AA}$ vs $N_{part}$ for $J/\psi$ production in Cu+Cu and Au+Au collisions at midrapidity with the prediction curves~\cite{Vogt2} with the EKS98 shadowing model~\cite{EKS}. 
      b) The same figure at forward rapidity. 
      c) Forward/mid rapidity $R_{AA}$ ratio. 
      d, e, f) The same figures with the predictions curves with the nDSg shadowing model~\cite{NDSG}. 
      The breakup cross sections for the prediction curves are obtained from fits to the $d+$Au data for each rapidity bin. 
      If the forward/mid rapidity $R_{AA}$ ratio is less than unity, it means rapidity narrowing. 
    }
    \label{fig3}
  \end{center}
\end{figure}

The final results of the $J/\psi$ measurement in Cu+Cu collisions are reported here and in~\cite{Adare4} with the published results in Au+Au collisions~\cite{Adare2}. 
The numbers of observed $J/\psi$ in Cu+Cu collisions at mid and forward rapidity are about 2050 and 9000, respectively. 
Compared to the Au+Au data, the Cu+Cu data provides a more precise $R_{AA}$ measurement for $N_{part}<100$, where the CNM effects might be dominant. 
Figure~\ref{fig2} shows $R_{AA}$ as a function of $p_T$ for each centrality bin in Cu+Cu collisions at mid and forward rapidity. 
Suppression by a factor of two is seen at both mid and forward rapidity in the most central collisions. 
No strong $p_T$ dependence of $R_{AA}$ is observed. 
Analysis is ongoing to extend $R_{AA}$ beyond $p_T$=5~GeV/$c$, where the STAR experiment has a moderate acceptance~\cite{Tang}.
In central Au+Au collisions, $R_{AA}$ slightly increases with $p_T$ at forward rapidity and $R_{AA}$ is flat within errors below $p_T$=5~GeV/$c$ at midrapidity. 

Figure~\ref{fig3} shows $R_{AA}$ at mid and forward rapidity and the forward/mid rapidity $R_{AA}$ ratio in Cu+Cu and Au+Au collisions. 
The Cu+Cu and Au+Au data agree well in the overlap region.  
The results of the $J/\psi$ measurement in $d+$Au collisions, which is the primary tool to investigate the CNM effects, were updated and are reported in~\cite{Adare3, Wysocki}. 
PHENIX has interpreted the $d$+Au data and extrapolated prediction of the CNM effects to heavy ion collisions based on calculations of R.~Vogt including nuclear shadowing and nuclear breakup~\cite{Vogt3, Vogt2}. 
While the nuclear breakup cross sections of $J/\psi$, $\sigma_{breakup}$, does not vary with rapidity in the framework of the calculations, PHENIX treated $\sigma_{breakup}$ as a rapidity dependent ``effective'' parameter and determined it by fitting to the $d$+Au data~\cite{Adare4}. 
The theoretical prediction curves with $\pm 1\sigma$ bands obtained for the rapidity dependent ``effective'' $\sigma_{breakup}$ with two nuclear shadowing models (EKS98~\cite{EKS} and nDSg~\cite{NDSG}) are also shown in figure~\ref{fig3}. 
Since the obtained $\sigma_{breakup}$ at forward rapidity is larger than one at midrapidity, the expected CNM effects from the PHENIX ``effective'' model at forward rapidity are larger than those at midrapidity and this leads to rapidity narrowing. 
The $J/\psi$ suppression beyond CNM effects seems to start at $N_{part}\sim200$ (100) at mid (forward) rapidity. 
As shown in figure~\ref{fig3} c) and f), rapidity narrowing in central Au+Au collisions is consistent with the expected CNM effects. 

While the current uncertainty of the CNM effects is large, it is expected to be reduced by the $d+$Au collision data collected in 2007--2008 whose statistics are 30 times larger than the currently available $d$+Au statistics.


\section*{References}
\begin{thebibliography}{15}
\bibitem{Matsui_Satz} Matsui~T and Satz~H 1986 {\it Phys. Lett.} B {\bf 178} 416 
\bibitem{Satz} Satz~H 2006 {\it J. Phys. G: Nucl. Part. Phys.} {\bf 32} R25 
\bibitem{Adare2} Adare~A {\it et al} (PHENIX Collaboration) 2007 {\it Phys. Rev. Lett.} {\bf 98} 232301 
\bibitem{Adare4} Adare~A {\it et al} (PHENIX Collaboration) 2008 {\it Preprint} arXiv:0801.0220 [nucl-ex] 
\bibitem{Silvestre} Silvestre~C (for the PHENIX Collaboration) 2008 {\it These Proceedings}, {\it Preprint} arXiv:0806.0475 [nucl-ex] 
\bibitem{Adare1} Adare~A {\it et al} (PHENIX Collaboration) 2007 {\it Phys. Rev. Lett.} {\bf 98} 232002 
\bibitem{PDG} Yao~W-M {\it et al} (Particle Data Group) 2006 {\it J. Phys. G: Nucl. Part. Phys.} {\bf 33} 1
\bibitem{Morino} Morino~Y (for the PHENIX Collaboration) 2008 {\it These Proceedings}, {\it Preprint} arXiv:0805.3871 [hep-ex] 
\bibitem{Vogt1} Vogt~R 2002 {\it Nucl. Phys.} A {\bf 700} 539 
\bibitem{Tang} Tang~Z (for the STAR Collaboration) 2008 {\it These Proceedings}, {\it Preprint} arXiv:0804.4846 [nucl-ex] 
\bibitem{Adare3} Adare~A {\it et al} (PHENIX Collaboration) 2008 {\it Phys. Rev.} C {\bf 77} 024912 
\bibitem{Wysocki} Wysocki~M (for the PHENIX Collaboration) 2008 {\it These Proceedings}, {\it Preprint} arXiv:0804.4660 [nucl-ex] 
\bibitem{Vogt3} Vogt~R 2005 {\it Phys. Rev.} C {\bf 71} 054902 
\bibitem{Vogt2} Vogt~R 2006 {\it Acta Phys. Hung.} A {\bf 25} 97 
\bibitem{EKS} Eskola~K~J, Kolhinen~V~J and Salgado~C~A 1999 {\it Eur. Phys. J.} C {\bf 9} 61
\bibitem{NDSG} de~Florian~D and Sassot~R 2004 {\it Phys. Rev.} D {\bf 69} 074028 
\end{thebibliography}
\end{document}